\documentclass[twoside]{article}
\newcommand{\bPf}{\par\vspace*{-4pt}\indent{\sc Proof.}\enskip}
\newcommand{\ePf}{\medskip}
\def\QED{\hskip0.1em\hfill\null\ \null\nobreak\hfill\kern3pt\vbox{\hrule\hbox
   {\vrule\kern1pt\vbox{\kern1.7pt\hbox{$\scriptscriptstyle{QED}$}
    \kern0.2pt}\kern1pt\vrule}\hrule}}

\def\END{\hskip0.1em\hfill\null\ \null\nobreak\hfill\kern3pt\vbox{\hrule\hbox
   {\vrule\kern1pt\vbox{\kern1.7pt\hbox{$\,\,\,\vspace{5pt}$}
    \kern0.2pt}\kern1pt\vrule}\hrule}}
\newtheorem{theorem}{Theorem}
\newtheorem{lemma}{Lemma}
\newtheorem{corollary}{Corollary}
\newtheorem{proposition}{Proposition}
\newtheorem{remark}{Remark}
\newtheorem{definition}{Definition}
\newtheorem{example}{Example}
\newcommand{\bCd}{\bEq\begin{CD}}
\newcommand{\eCd}{\end{CD}\eEq}
\newcommand{\bcd}{\beq\begin{CD}}
\newcommand{\ecd}{\end{CD}\eeq}
\newcommand{\ben}{\begin{enumerate}}
\newcommand{\een}{\end{enumerate}}
\newcommand{\bEq}{\begin{eqnarray}}
\newcommand{\eEq}{\end{eqnarray}}
\newcommand{\beq}{\begin{eqnarray*}}
\newcommand{\eeq}{\end{eqnarray*}}
\newcommand{\bDf}{\begin{definition}\em}
\newcommand{\eDf}{\end{definition}}
\newcommand{\bLm}{\begin{lemma}}
\newcommand{\eLm}{\end{lemma}}
\newcommand{\bPr}{\begin{proposition}}
\newcommand{\ePr}{\end{proposition}}
\newcommand{\bTh}{\begin{theorem}}
\newcommand{\eTh}{\end{theorem}}
\newcommand{\bCr}{\begin{corollary}}
\newcommand{\eCr}{\end{corollary}}
\newcommand{\bRm}{\begin{remark}\em}
\newcommand{\eRm}{\end{remark}}
\newcommand{\bEx}{\begin{example}\em}
\newcommand{\eEx}{\end{example}}

\newcommand{\ie}{{\em i.e$.$} }
\newcommand{\eg}{{\em e.g$.$} }
\newcommand{\R}{I\!\!R}

\newcommand{\der}{\partial}

\newcommand{\cH}{\mathcal{H}}

\newcommand{\sub}{\subset}

\newcommand{\wed}{\wedge}
\newcommand{\com}{\!\circ\!}
\newcommand{\ten}{\!\otimes\!}

\newcommand{\alp}{\alpha}

\newcommand{\gam}{\gamma}

\newcommand{\lam}{\lambda}
\newcommand{\sig}{\sigma}

\newcommand{\ome}{\omega}
\newcommand{\Gam}{\Gamma}
\newcommand{\Del}{\Delta}

\newcommand{\Ome}{\Omega}

\newcommand{\vartht}{\vartheta}

\newcommand{\Tht}{\Theta}
\newcommand{\For}{{\Lambda}}

\title{\large{{\bf Some aspects of the homogeneous formalism in Field Theory and gauge
invariance}\thanks{This paper is in final form and will not be submitted elsewhere.}}}
\author{{\normalsize M. Palese and E. Winterroth\thanks{Both authors are supported by GNFM of
INdAM, University of Torino and MIUR (PRIN 2005). The second author (E. W.) is also supported by a
PhD grant of the University of Torino - {\em Dottorato di Ricerca in Matematica XVIII Ciclo}.}}
\\{\footnotesize Department of Mathematics,
University of Torino}
\\{\footnotesize via C. Alberto 10, 10123 Torino, Italy}\\ 
{\footnotesize e--mails: 
{\sc marcella.palese@unito.it, ekkehart@dm.unito.it}}}
\date{}
\begin{document}

\maketitle

\begin{abstract}

We propose a suitable formulation of the Hamiltonian formalism for Field Theory in
terms of Hamiltonian connections and multisymplectic 
forms where a 
composite fibered bundle, involving a line bundle, plays the role of 
an extended configuration bundle. This new approach can be interpreted as 
a suitable generalization to Field Theory of the homogeneous formalism
for Hamiltonian Mechanics. As an example of application, we obtain the expression of a {\em formal
energy} for a parametrized  version of
the Hilbert--Einstein Lagrangian and we show that this quantity is conserved.

\medskip



\end{abstract}

\section{Introduction}

A geometric setting for the Hamiltonian description of Field Theory is 
proposed which generalizes 
the homogeneus Hamiltonian formalism in time-dependent Mechanics (see \eg \cite{Sar98}). 
The aim is to provide a suitable description of the gauge character appearing in the covariant
formulations of Hamiltonian {\em multiphase} Field Theory and their
quantizations  based on the seminal paper by Dedecker \cite{Ded77} and developed in the recent
literature by many authors; see \eg
\cite{Cac1,Cac2,Kan98,Kan01,Krup01,RRS04,RRS06} and many references quoted therein.

One of the main features of our approach is that one can describe the polymomenta and other 
objects such as Hamiltonian forms in terms of differential forms with values in the vertical
tangent bundle of an appropriate line bundle $\Tht$. 
The introduction of the line bundle $\Tht$, in fact, can be understood as a suitable way
of describing the {\em gauge} character appearing in the multiphase formalism 
for Field Theory, essentially due to the fact that the independent variables are more than one   
and thus the Poincar\'e--Cartan invariant is defined only up to the choice of a
symmetric linear connection on the basis manifold (see \eg
\cite{GoSt73,Kij74,Kol83,Krup01}). 

With the aim of overcoming this ambiguities, instead of bundles over an
$n$--dimensional base manifold $X$, we consider {\em fibrations over a line bundle $\Tht$
fibered over $X$}. We recall that a geometric formulation of the Hamiltonian formalism for Field 
Theory in
terms of Hamiltonian connections and multisymplectic forms was developed \eg in
\cite{MaSa00,Sar95,Sar98}. In this framework, the covariant Hamilton
equations for Mechanics and Field Theory are defined in terms of multisymplectic $(n+2)$--forms,
where $n$ is the dimension of the basis manifold, together with connections on the 
configuration bundle. Following the analogous setting for Mechanics and
for the polymomentum approach to Field Theory, we propose a new
concept of event bundle, configuration bundle and Legendre bundle. Correspondingly, 
Hamiltonian connections,
Hamiltonian forms and covariant Hamilton equations can be suitably described in this
framework. This new approach takes into account the existence of more than one independent variable
in Field Theory, but enables us to keep most of the features of
time-dependent Hamiltonian Mechanics. 
In fact, the prominent role of {\em symplectic} structures in field theories has been
stressed in 
\cite{Kij73,Kij74,KiSz76,KiTu79} and recently a symplectic approach for the study of 
Canonical Gravity \cite{Kij01} has been proposed. 

We point out that  
the extension of the Hamiltonian formalism from Mechanics to Field Theory is usually
performed starting from the {\em non-homogeneous} formalism of Mechanics, where a gauge 
choice is assumed {\em a priori} to be performed; precisely $q^{0}(t)=t$, where $t$ is the time. 
It is however well known - and it deserves to be noticed within
our note - that Mechanics is invariant with respect to gauge choices of this kind, \ie with
respect to choices of the section $q^{0}(t)$; see
\eg the review in \cite{Kij01}. 
Accordingly with the just mentioned approach to Mechanics, in Hamiltonian Field Theory the
configuration variables (fields) are usually assumed to depend {\em directly} on a number of
independent variables greater than one.   As an outcome, it is well known
that  polymomenta correspondingly defined are in a bigger number than the configurations, and that
the corresponding  Hamiltonians
do not have a clear interpretation as physical observables. Many difficulties 
arises in the
attempts of quantization of such a Hamiltonian Field Theory (see \eg the detailed reviews in
\cite{HeKo04a,HeKo04b,Kan01}).  

In the present paper we generalize to Field Theory the so-called {\em homogeneous formalism 
of time-dependent Mechanics}, so that a local line
coordinate $\tau$ plays the role of the `homogeneous' local coordinate $q^{0}$, and a {\em formal
Hamiltonian theory} is constructed where the Hamiltonian describes the {\em formal evolution} along
the line coordinate. The latter in turn depends on the basis (independent) coordinates
{\em when a gauge choice is performed}, \ie a section of the line bundle is chosen. This is nearer
to \cite{CrWi87}, and it keeps the advantages of a finite dimensional approach. 
Thus the formal Hamiltonian can be interpreted as a formal energy 
(beeing the conjugated momentum to the formal
evolution parameter). The energy then is a gauge charge since it
is related with invariance properties with respect to infinitesimal transformations of the line
(vertical) coordinate.

In Section \ref{2} we state the general framework of composite fiber bundles, their jet
prolongations and composite connections. Section \ref{3} contains {\em abstract Hamilton equations} 
and a
Theorem which relates the {\em abstract Hamiltonian dynamics} introduced  here with the
standard Hamilton--De Donder equations (see \cite{Krup01} for a detailed review on the topic and
recent developments).
Proceeding in analogy with Mechanics we obtain the expression of a
{\em `formal' energy} for an extended version of the Hilbert--Einstein Lagrangian and 
we show that this quantity is conserved.

The present approach is 
a completion of \cite{FPW01} where
the formal aspect of the homogeneus setting was not exhaustively explicated. Some misprints and
imprecisions there appearing will be also corrected.

\section{Jets and connections on composite bundles}\label{2}

The general framework is a fibered bundle over $X$, $\pi : Y \to X$,
with $\textstyle{dim} X = n$ and $\textstyle{dim} Y = n+m$ 
and, for $r \geq 0$, its jet manifold $J_r Y$. We recall the natural fiber 
bundles
$\pi^r_s : J_r Y \to J_s Y$, $r \geq s$, $\pi^r : J_r Y \to X$, and, 
among these, the {\em affine\/} fiber bundles $\pi^r_{r-1}$.

Greek indices $\lam ,\mu ,\dots$ run from $1$ to $n$ and they label base 
coordinates, while
Latin indices $i,j,\dots$ run from $1$ to $m$ and label fibre coordinates,
unless otherwise specified.
We denote multi--indices of dimension $n$ by underlined Greek letters such as
$\underline{\alpha} = (\alp_1, \dots, \alp_n)$, with $0 \leq \alp_\mu$, 
$\mu=1,\ldots,n$; by an abuse 
of notation, we denote with $\lam$ the multi--index such that 
$\alp_{\mu}=0$, if $\mu\neq \lam$, $\alp_{\mu}= 1$, if 
$\mu=\lam$.
We also set $|\underline{\alpha}| \doteq \alp_{1} + \dots + \alp_{n}$.
The charts induced on $J_r Y$ are denoted by $(x^\lam,y^i_{\underline{\alpha}})$, with $0
\leq |\underline{\alpha}| \leq r$; in particular, we set $y^i_{\bf{0}}
\equiv y^i$. The local bases of vector fields and $1$--forms on $J_r Y$ induced by
the coordinates above are denoted by $(\der_\lam ,\der^{\underline{\alpha}}_i)$ and 
$(d^\lam,d^i_{\underline{\alpha}})$,
respectively.

For $ r\geq 1$, the {\em contact maps\/} on jet spaces induce
the natural complementary fibered
morphisms over the affine fiber bundle $J_r Y \to J_{r-1}Y$
\begin{equation}\label{affine1}
D_{r} : J_r Y \times_{X} TX \to TJ_{r-1}Y \,,
\quad 
\vartht_{r} : J_r Y \times_{J_{r-1}Y} TJ_{r-1}Y \to VJ_{r-1}Y \,,
\end{equation}
with coordinate expressions, for $0 \leq |\underline{\alpha}| \leq r-1$, given by
$D_{r} = d^\lam\ten {D}_\lam = d^\lam\ten
(\der_\lam + y^j_{\underline{\alpha}+\lam}\der_j^{\underline{\alpha}})$, 
$\vartht_{r} = \vartht^j_{\underline{\alpha}}\ten\der_j^{\underline{\alpha}} =
(d^j_{\underline{\alpha}}-y^j_{{\underline{\alpha}}+\lam}d^\lam)
\ten\der_j^{\underline{\alpha}}$,
and the natural fibered splitting $J_r Y\times_{J_{r-1}Y}T^*J_{r-1}Y$ $=$
$J_r Y\times_{J_{r-1}Y}\left(T^*X \oplus V^{*}J_{r-1}Y\right)$.

\begin{definition}
A {\em connection} on the fiber bundle $Y \to X$ is defined by the (mutually dual) linear bundle
morphisms over $Y$: $Y\times_{X}TX \to TY$, $V^{*}Y\to T^{*}Y$
which split the exact sequences 
\beq
0 \to VY \hookrightarrow TY \to Y\times_{X}TX \to 0 \,,
\quad
0 \to Y\times_{X}T^{*}X \hookrightarrow T^{*}Y \to V^{*}Y \to 0 \,.
\eeq\END
\end{definition}

We recall that there is a one--to--one correspondence between the connections $\Gam$ on a fiber bundle
$Y\to X$ and the global sections 
\,$\Gam:Y\to J_{1}Y$\,
of the affine jet bundle $J_{1}Y\to Y$ (see \eg \cite{MaSa00}).

In the following a relevant role is played by the composition of fiber bundles
\begin{equation}\label{compositebundle}
Y\to\Tht \to X\,,
\end{equation}
where $\pi_{YX}: Y\to X$, $\pi_{Y\Tht}:Y\to\Tht$ and $\pi_{\Tht X}:
\Tht\to X$ are fiber bundles. The above composition was introduced under the name of {\em
composite fiber bundle} in
\cite{MaSa98,Sar95} and shown to be useful for physical applications, \eg for the
description of mechanical systems with time--dependent parameters.

\medskip

We shall be concerned here with the description of connections on
composite fiber bundles. We will follow the notation and main results stated in \cite{MaSa00};
see also \cite{CaKo93}.

We shall denote by $J_{1}\Tht$, $J_{1}^{\Tht}Y$ and $J_{1}Y$, the jet
manifolds of the fiber bundles $\Tht\to X$, $Y\to\Tht$ and $Y\to X$ 
respectively.

Let $\gam$ be a connection on the composite bundle $\pi_{YX}$ projectable over a
connection $\Gam$ on $\pi_{\Tht X}$, \ie such
$J_{1}\pi_{Y\Tht}\,\com\,\gam\,=\,\Gam\,\com\,\pi_{Y\Tht}$.
Let $\gamma_{\Tht}$ be a connection on the fiber bundle $\pi_{Y\Tht}$. 
Given a
connection $\Gam$ on $\pi_{\Tht X}$, there exists \cite{MaSa00} a canonical morphism over $Y$,
$\rho: J_{1}\Tht\times_{X}J_{1}^{\Tht}Y\to J_{1}Y$, which sends
$(\Gam,\gamma_{\Tht})$, into the {\em composite connection} $\gam \doteq
\gamma_{\Tht}\com \Gam$ on
$\pi_{YX}$, projectable over $\Gam$.
Recall that given a composite fiber bundle (\ref{compositebundle}) and a global section $h$ of the
fiber bundle $\pi_{\Tht X}$, then the restriction $Y_{h}\doteq h^{*}Y$ of the fiber bundle
$\pi_{Y\Tht}$ to $h(X)\sub \Tht$ is a subbundle $i_{h}:Y_{h}\hookrightarrow Y$ of the
fiber bundle $Y\to X$ \cite{MaSa00}.
Let then $h$ be a section of $\pi_{\Tht X}$. Every connection
$\gamma_{\Tht}$ induces the pull--back connection $\gamma_{h}$ on the subbundle
$Y_{h}\to X$. The composite connection
$\gam\,=\,\gamma_{\Tht}\,\com\,\Gam$ is reducible to $\gamma_{h}$ if and only if
$h$ is an integral section of $\Gam$.

\medskip 

We have the following exact sequences of {\em vector bundles
over a composite bundle $Y$}:
\begin{equation}
0\to V_{\Tht}Y\hookrightarrow VY\to Y\times_{\Tht}V\Tht\to 0\,, \qquad 
0\to Y\times_{\Tht}V^{*}\Tht \hookrightarrow V^{*}Y\to V^{*}_{\Tht}Y \to 0\,,
\end{equation}
where $V_{\Tht}Y$ and $V^{*}_{\Tht}Y$ are the vertical tangent and cotangent bundles to the
bundle
$\pi_{Y\Tht}$.

\begin{remark}
Every connection $\gamma_{\Tht}$ on $\pi_{Y\Tht}$ provides the dual splittings
\begin{equation}
VY = V_{\Tht}Y\oplus_{Y}\gamma_{\Tht}(Y\times_{\Tht}V\Tht)\,, \qquad
V^{*}Y = Y\times_{\Tht}V^{*}\Tht \oplus_{Y}\gamma_{\Tht}(V^{*}_{\Tht}Y)\,,
\end{equation}
of the above exact sequences.
By means of these splittings one can easily construct the {\em vertical covariant differential} on the
composite bundle $\pi_{YX}$, \ie a first order differential operator 
\begin{equation}\label{verdiff}
{\Del} _{\gamma_{\Tht}}: J_{1}Y \to T^{*}X
\oplus_{Y} V^{*}_{\Tht}Y\,.
\end{equation}
This operator is characterized by the property that the restriction of $\Del_{\gamma_{\Tht}}$, induced by a section $h$ of $\pi_{\Tht X}$, coincides
with the covariant differential on
$Y_{h}$ relative to the pull--back connection $\gamma_{h}$ \cite{MaSa00}.\END
\end{remark}

\section{Homogeneus formalism in Field Theory}\label{3}

We recall now that the covariant Hamiltonian Field Theory can be conveniently formulated in terms
of Hamiltonian connections and Hamiltonian forms \cite{Sar95}. Here we shall construct a
Hamiltonian formalism for Field Theory as a theory on the composite {\em event bundle} $Y \to\Tht\to
X$, with $\pi_{\Tht X}: \Tht \to X$ a {\em line bundle} having local fibered coordinates $(x^{\lam},
\tau)$.

Let us now consider the {\em extended homogeneous Legendre bundle} $Z_{Y}\doteq
T^{*}Y\wed (\For^{n}T^{*}\Tht)\to X$. It is the trivial one-dimensional bundle $\kappa: Z_{Y} \to
\Pi_{\Tht}$, where $\Pi_{\Tht}\doteq V^{*}Y\wed (\For^{n}T^{*}\Tht)\to X$ is {\em extended 
Legendre bundle}.
There exists a canonical isomorphism
\begin{equation}
\Pi_{\Tht}\simeq \For^{n+1}T^{*}\Tht\ten_{Y}V^{*}Y\ten_{Y}T\Tht\,.
\end{equation}

\begin{definition}
We call the fiber bundle $\pi_{Y\Tht}: Y\to\Tht$ the {\em abstract event space} of the field
theory. The {\em configuration space} of the Field Theory is then the first order jet manifold
$J^{\Tht}_{1}Y$. 
The {\em abstract Legendre bundle} of the field
theory is the fiber bundle
$\Pi_{\Tht}\to\Tht$.
\END\end{definition}

Let now $\gamma_\Tht$ be a connection on $\pi_{Y\Tht}$ and $\Gam_{\Tht}$ be a
connection on
$\pi_{\Tht X}$. We have the following non--canonical isomorphism
\begin{equation}\label{iso}
\Pi_{\Tht}\simeq_{(\gamma_\Tht,\Gam_{\Tht})}
\For^{n+1}T^{*}\Tht\ten_{Y}[(Y\oplus_{\Tht}V^{*}\Tht)\oplus_{Y}
\gamma_{\Tht}(V^{*}_{\Tht}Y)]\ten_{Y}(V\Tht \oplus_{\Tht}H\Tht)\,.
\end{equation}

In this perspective, we consider the canonical bundle monomorphism over $Y$ providing the
tangent--valued Liouville form on $\Pi_{\Tht}$, \ie
\begin{equation}
\vartht_{Y}: \Pi_{\Tht}\hookrightarrow \For^{n+2}T^{*}Y\ten_{Y}(V\Tht \oplus_{\Tht}
H\Tht)\,,
\end{equation}
where $H\Tht$ is the horizontal subbundle.

\medskip

Let now $(x^{\hat{\mu}})=(x^{\mu}, \tau)$, $\hat{\ome}=d^{\mu_1}\wed d^{\mu_2}\wed \ldots\wed
d^{\mu_n}\wed d\tau$, $\der_{\hat{\mu}}=(\der_{\mu},\der_{\tau})$ be, respectively,  local
coordinates on $\Tht$, the induced volume form, local generators of tangent vector fields and put 
$\hat{\ome}_{\hat{\lam}}\doteq \der_{\hat{\lam}}\rfloor \hat{\ome}$.

Inspired by \cite{KiTu79} we set 
$\bar{p}_i \doteq p^{\hat{\mu}}_{i}\ten \der_{\hat{\mu}}$ and obtain
\begin{equation}
\vartht_{Y}=\bar{p}_{i}d^{i}\wed\hat{\ome}\,.
\end{equation}

The polysymplectic form $\Ome_{Y}$ on $\Pi_{\Tht}$ is then intrinsically defined by
$\Ome_{Y}\rfloor\psi=d(\vartht_{Y}\rfloor\psi)$,
where $\psi$ is an arbitrary $1$--form on $\Tht$; its coordinate expression is given
by 
\bEq\label{multisympl}
\Ome_{Y}=
d\bar{p}_{i}\wed d^{i}\wed\hat{\ome} \simeq d\bar{p}_{i}\wed d^{i}\wed d\tau\,.
\eEq

\medskip 

Let $J_{1}\Pi_{\Tht}$ be the first order jet manifold of the extended Legendre
bundle
$\Pi_{\Tht}\to X$. A connection $\gam$ on the extended Legendre bundle is then in
one--to--one correspondence with a global section of the affine bundle $J_{1}
\Pi_{\Tht}\to \Pi_{\Tht}$.
Such a connection is said to be a {\em Hamiltonian
connection} iff the exterior form
$\gam\rfloor\Ome_{Y}$ is closed.

A Hamiltonian $\cH$ on $\Pi_{\Tht}$ is defined as a section $\bar{p} =-\cH$ of the bundle $\kappa$.
Let $\gam$ be a Hamiltonian connection on $\Pi_{\Tht}$ and $ U$ be an open subset of $\Pi_{\Tht}$.
Locally, we have $\gam\rfloor\Ome_{Y} = (d\bar{p}_{i}\wed d^i - d{\cal H}) \wed
\hat{\ome}\simeq (d\bar{p}_{i}\wed d^i - d{\cal H}) \wed
d\tau \doteq dH$, where ${\cal H}:  U\sub\Pi_{\Tht} \to V\Tht$ and
$d=\der_{i}d^{i}+\bar{\der}^{i}d\bar{p}_{i}+\der_{\tau}d\tau$ is the total differential on
$V^{\Tht}\Pi_{\Tht}$.

The local mapping ${\cal H}:  U\sub\Pi_{\Tht} \to V\Tht$ is called a {\em Hamiltonian}. 
The form
$H$ on the extended Legendre bundle $\Pi_{\Tht}$ is called a {\em Hamiltonian form}. 
Every Hamiltonian form $H$ admits a Hamiltonian connection $\gam_{H}$ such that the following holds:
$\gam_{H}\rfloor\Ome_{Y}= dH$.

We define the {\em abstract covariant Hamilton equations} to be the 
kernel of the first order differential operator ${\Del} _{{\tilde{\gamma}}_{\Tht}}$ defined as the vertical
covariant differential (see Eq. \ref{verdiff}) relative to the connection $\tilde{\gamma}_{\Tht}$ on 
the abstract Legendre bundle $\Pi_{\Tht}\to\Tht$.

In the following a `dot' stands for $\pounds_{\der_{\tau}}$, \ie the Lie derivative along $\der_{\tau}$.
In this case the Hamiltonian form $H$ is the
Poincar\'e--Cartan form of the {\em  Lagrangian}
$L_{H}=(\bar{p}_{i}\dot{y}^{i} -{\cal H})\,d\tau$ on $V^{\Tht} \Pi_{\Tht}$, with values in $V\Tht$.
Furthermore, the {\em Hamilton operator} for
$H$ is defined as the Euler--Lagrange operator associated with
$L_{H}$, namely ${\cal E}_{H}:$ $V^{\Tht} \Pi_{\Tht}$ $\to$ $T^{*}\Pi_{\Tht}\wed\For^{n+1}T^{*}X$.

\medskip
We state then  the following.

\bPr
The kernel of the Hamilton operator, \ie the Euler--Lagrange equations for $L_{H}$, is
an affine closed embedded subbundle of $V^{\Tht}\Pi_{\Tht}\to \Pi_{\Tht}$, locally given by the
{\em covariant formal Hamilton equations} on
the extended Legendre bundle $\Pi_{\Tht}\to X$
\begin{equation}
\dot{y}^{i}= \bar{\der}^{i}{\cal H}\,,
\end{equation}
\begin{equation}
\dot{\bar{p}}_{i} =-\der_{i} {\cal H}\,,
\end{equation}
\begin{equation}\label{conservation}
\dot{{\cal H}}=\der_{\tau} {\cal H}\,.
\end{equation}
\ePr

\medskip

These latter results could be compared with 
\cite[Sec.$4$]{Kan98}. However, within the limits of the purpose of this note, in the following we
just recall  the relation with the standard polysymplectic approach 
(for a review of the  topic see \eg
\cite{Ded77,Kan98,Kij73,KiTu79,Krup01} and references quoted 
therein) and provide an example of application. 

\begin{lemma}\label{main}
Let $\gam_{H}$ be a Hamiltonian connection on $\Pi_{\Tht}\to X$. 
Let $\tilde{\gamma}_{\Tht}$ and $\Gam$ be connections on 
$\Pi_{\Tht}\to Y$ 
and $\Tht\to X$, respectively. Let $\sig$ and $h$ be sections of the 
bundles $\pi_{Y\Tht}$ and $\pi_{\Tht X}$, respectively.

Then the standard Hamiltonian connection on $\Pi_{\Tht}\to X$ turns out to
be the pull--back connection $\tilde{\gamma}_{\phi}$ induced on the subbundle
$\Pi_{\Tht\,\phi}\hookrightarrow \Pi_{\Tht}\to X$ by the section $\phi=h\com 
\sig$ of $Y\to X$.
\end{lemma}

\bPf
The abstract Legendre bundle is in fact a composite bundle 
$\Pi_{\Tht}\to Y\to \Tht$. Our claim then 
follows for any section $\phi$ of the composite bundle 
$Y\to\Tht\to X$ of the type $\phi=h\com \sig$, since the extended 
Legendre bundle $\Pi_{\Tht}\to X$ can be also seen as the composite 
bundle $\Pi_{\Tht}\to Y\to X$.
\QED\ePf

As a straightforward consequence  we can state the following \cite{FPW01}

\bPr\label{final}
Let ${\Del} _{\tilde{\gamma}\,, \phi}$ be the covariant differential on the subbundle
$\Pi_{\Tht\,\phi}\hookrightarrow \Pi_{\Tht}\to X$ relative to the 
pull--back connection $\tilde{\gamma}_{\phi}$. The kernel of 
${\Del} _{\tilde{\gamma}\,, \phi}$ coincides with the Hamilton--De Donder 
equations of the standard polysymplectic approach to field theories.
\ePr

\medskip

\bEx ({\bf Formal gravitational energy.})

Let us now specify the above formalism for an extended version of the Hilbert--Einstein Lagrangian, 
\ie essentially
the Hilbert--Einstein Lagrangian for a metric $g$ parametrized by the line coordinate $\tau$.

Let then
$\textstyle{dim}X = 4$ and
$X$ be orientable. 
Consider the configuration composite bundle $Lor(X)_{\Tht}\to \Tht\to X$ coordinated by
$(g^{\mu\nu},\tau,x^{\lam})$, where $(\tau,x^{\lam})$ are coordinates of the line bundle $\Tht$ and
$g^{\mu\nu}$ are Lorentzian metrics on $X$ (provided that they exist), \ie sections of
$Lor(X)\to X$. We call $Lor(X)_{\Tht}$ the bundle of Lorentzian metrics (on $X$) {\em
parametrized by $\tau$}. 
The bundle $Lor(X)_{\Tht}\to \Tht$ is not necessarily trivial; it is characterized as follows. Every
section
$h$ of the line bundle  $\Tht$ defines  the restriction $h^{*}Lor(X)_{\Tht}$ of $Lor(X)_{\Tht}\to
\Tht$ to
$h(X)\sub \Tht$, which is a subbundle of $Lor(X)_{\Tht}\to X$. One can think of
$h^{*}Lor(X)_{\Tht}\to X$ as being the bundle $Lor(X)$ of Lorentzian metrics on $X$ with
the background parameter function $h(x^{\mu})$ (similar considerations can be found in
parametrized Mechanics, see \eg \cite{MaSa00}). However in what follows we will not fix such a
section.

The {\em extended} Hilbert--Einstein Lagrangian is the form $\lam_{HE} =
L_{HE}\,\hat{\ome}\,$, were 
$L_{HE}=r\,\sqrt{\underline{g}}$. Here $r: J_{2}^{\Tht}(Lor(X)_{\Tht}) \to \R$ is the function 
such that, for any parametrized Lorentz metric $g$, we have $r\circ j^{\Tht}_2g = s$, 
being $s$ the scalar curvature associated with $g$, and $\underline{g}$ is the 
determinant of $g$. 

Put $\pi^{\mu\nu}=\sqrt{\underline{g}} g^{\mu\nu}$ and
$\phi^{\hat{\rho}\hat{\lam}}_{\mu\nu}
\doteq\der L_{HE}/ \der{\dot\pi}^{\mu\nu}_{\hat{\rho}\hat{\lam}}$.

Now, consider that
\begin{equation}
L_{HE}=\pi^{\mu\nu}\hat{R}_{\mu\gam}=\bar{\phi}_{\mu\nu}{\dot\pi}^{\mu\nu}-{\cal H}\,,
\end{equation}
where $\hat{R}_{\mu\gam}=\hat{R}^{\lam}_{\mu\lam\gam}$ denotes the components of the Ricci tensor of
the Lie-dragged metric and $\bar{\phi}_{\mu\nu}\equiv \phi^{\hat{\rho}\hat{\lam}}_{\mu\nu}\ten
\der_{\hat{\rho}\hat{\lam}}$. Hence the {\em formal Hamiltonian} turns out to be
\begin{equation}
{\cal H}=(-\pi^{\mu\gam}\hat{R}_{\mu\gam}+\bar{\phi}_{\mu\nu}{\dot\pi}^{\mu\nu})\,.
\end{equation}

Notice that the formal Hamiltonian does not depend explicitly on $\tau$. From 
the covariant Hamilton equations, in particular from Eq. (\ref{conservation}), we have
$\dot{{\cal H}}=0$;
thus the formal Hamiltonian turns out to be a
conserved quantity. In fact we can interprete it as a conserved formal energy for the gravitational
field (compare with \cite{Kij01} where an analogous approach is followed by defining the Cauchy
data on a three-dimensional submanifold of space-time).\END
\eEx

We stress that, as far as a section
$h(x^{\mu})$  of $\Tht\to X$ has not been fixed {\em a priori}, our approach provides an appropriate
covariant Hamiltonian description of gravitation, which does not require neither a
$(3+1)$ splitting of space-time - as it is done in the ADM-like formalisms - nor the fixing of a
background connection - as it is done whitin the Palatini-like approaches. Both of the latter
approaches we mentioned, in fact, can provide Hamiltonian descriptions of gravitation, which however
loose the required genuine covariance.

\medskip

We finally notice that this formal approach stresses
the underlying algebraic structure of Field Theory, which was shown to be related
with a new
$K$--theory for vector bundles carrying the same kind of `special' multisymplectic structure
\cite{Win01} (related multisymplectic $3$-forms on manifolds
have been also studied \eg in \cite{BuVa04,PaVa03}). 


\end{document}